\documentclass{aastex}
\usepackage{emulateapj5}
\usepackage{apjfonts}
\usepackage{graphicx}

\def\HETE{{\it HETE }}
\def\MO{{\it Mars Odyssey }}
\def\Ulysses{{\it Ulysses }}
\def\BeppoSAX{{\it BeppoSAX }}
\def\CGRO{{\it Compton Gamma-Ray Observatory}}

\shorttitle{HETE-2 Observations of GRB020531}
\shortauthors{Lamb et al.}

\begin{document}
\title{HETE-2 Localization and Observations of the Short, Hard Gamma-Ray
Burst GRB020531}

\author{
D. Q. Lamb,\altaffilmark{1}
G. R. Ricker,\altaffilmark{2}
J.-L. Atteia,\altaffilmark{3}
K. Hurley,\altaffilmark{4}
N. Kawai,\altaffilmark{5,6}
Y. Shirasaki,\altaffilmark{5,9}
T. Sakamoto,\altaffilmark{5,6,10}
T. Tamagawa,\altaffilmark{6}
C.~Graziani,\altaffilmark{1}
J.-F.~Olive,\altaffilmark{3}
A. Yoshida,\altaffilmark{5,7}
M.~Matsuoka,\altaffilmark{8}
K. Torii,\altaffilmark{6}
E.~E.~Fenimore,\altaffilmark{10}
M.~Galassi,\altaffilmark{10}
T.~Tavenner,\altaffilmark{10}
T.~Q.~Donaghy,\altaffilmark{1}
M. Boer,\altaffilmark{3}
J.-P.~Dezalay,\altaffilmark{3}
R.~Vanderspek,\altaffilmark{2}
G.~Crew,\altaffilmark{2}
J.~Doty,\altaffilmark{2}
G.~Monnelly,\altaffilmark{2}
J.~Villasenor,\altaffilmark{2}
N. Butler,\altaffilmark{2}
J.~G.~Jernigan,\altaffilmark{4}
A. Levine,\altaffilmark{2}
F. Martel,\altaffilmark{2} 
E. Morgan,\altaffilmark{2}
G. Prigozhin,\altaffilmark{2}
S. E. Woosley,\altaffilmark{11}
T.~Cline,\altaffilmark{12}
I~ Mitrofanov,\altaffilmark{13}
D.~Anfimov,\altaffilmark{13}
A. Kozyrev,\altaffilmark{13}
M. Litvak,\altaffilmark{13}
A. Sanin,\altaffilmark{13}
W.~Boynton,\altaffilmark{14}
C.~Fellows,\altaffilmark{14}
K.~Harshman,\altaffilmark{14}
C. Shinohara,\altaffilmark{14}
R. Starr,\altaffilmark{12}
J. Braga,\altaffilmark{15}
R.~Manchanda,\altaffilmark{16}
G.~Pizzichini,\altaffilmark{17}
K.~Takagishi,\altaffilmark{18}
 and M. Yamauchi\altaffilmark{18}
}

\altaffiltext{1}{Department of Astronomy and Astrophysics, University
of Chicago, 5640 South Ellis Avenue, Chicago, IL 60637.}

\altaffiltext{2}{Center for Space Research, Massachusetts Institute of
Technology, 70 Vassar Street, Cambridge, MA, 02139.}

\altaffiltext{3}{Centre d'Etude Spatiale des Rayonnements, CNRS/UPS,
B.P.4346, 31028 Toulouse Cedex 4, France.}

\altaffiltext{4}{Space Sciences Laboratory, University of California at
Berkeley, 601 Campbell Hall, Berkeley, CA, 94720.}

\altaffiltext{5}{Department of Physics, Tokyo Institute of Technology, 
2-12-1 Ookayama, Meguro-ku, Tokyo 152-8551, Japan.}

\altaffiltext{6}{RIKEN (Institute of Physical and Chemical Research),
2-1 Hirosawa, Wako, Saitama 351-0198, Japan.}

\altaffiltext{7}{Department of Physics, Aoyama Gakuin University,
Chitosedai 6-16-1 Setagaya-ku, Tokyo 157-8572, Japan.}

\altaffiltext{8}{Tsukuba Space Center, National Space Development
Agency of Japan, Tsukuba, Ibaraki, 305-8505, Japan.}

\altaffiltext{9}{National Astronomical Observatory, Osawa 2-21-1,
Mitaka,  Tokyo 181-8588 Japan.}

\altaffiltext{10}{Los Alamos National Laboratory, P.O. Box 1663, Los 
Alamos, NM, 87545.}

\altaffiltext{11}{Department of Astronomy and Astrophysics, University 
of California at Santa Cruz, 477 Clark Kerr Hall, Santa Cruz, CA
95064.}

\altaffiltext{12}{NASA Goddard Space Flight Center, Greenbelt, MD,
20771.}

\altaffiltext{13}{Space Research Institute, Profsojuznaya Str. 84/32,
117810, Moscow, Russia.}

\altaffiltext{14}{Department of Planetary Sciences, Lunar and Planetary
Laboratory, Tucson, AZ 85721-0092.}

\altaffiltext{15}{Instituto Nacional de Pesquisas Espaciais, Avenida
Dos Astronautas 1758, S\~ao Jos\'e dos Campos 12227-010, Brazil.}

\altaffiltext{16}{Department of Astaronomy and Astrophysics, Tata 
Institute of Fundamental Research, Homi Bhabha Road, Mumbai, 400 005, 
India.}

\altaffiltext{17}{Consiglio Nazionale delle Ricerche (IASF), via Piero
Gobetti, 101-40129 Bologna, Italy.}

\altaffiltext{18}{Faculty of engineeering, Miyazaki University, Gakuen
Kibanadai Nishi, Miyazaki 889-2192, Japan.}

\begin{abstract}
The {\it HETE-2} (hereafter \HETE) French Gamma Telescope (FREGATE) and
the Wide-field X-ray Monitor (WXM) instruments detected a short
($t_{50} = 360$ msec in the FREGATE 85-300 keV energy band),
hard gamma-ray burst (GRB) that occurred at 1578.72 SOD (00:26:18.72
UT) on 31 May 2002.  The WXM flight localization software produced a
valid location in spacecraft (relative) coordinates.  However, since no
on-board real-time star camera aspect was available, an absolute
localization could not be disseminated.  A preliminary localization was
reported as a GCN Position Notice at 01:54:22 UT, 88 min after the
burst.  Further ground analysis produced a refined localization, which
can be expressed as a 90\% confidence rectangle that is 67 arcminutes
in RA and 43 arcminutes in Dec (90\% confidence region), centered at RA
= +15$^{\rm h}$ 14$^{\rm m}$ 45$^{\rm s}$, Dec = -19$^\circ$ 21\arcmin
35\arcsec (J2000).  An IPN localization  of the burst was disseminated
18 hours after the GRB (Hurley et al. 2002b).  A refined IPN
localization was disseminated $\approx$ 5 days after the burst.  This
hexagonal-shaped localization error region is centered on RA = 15$^{\rm
h}$ 15$^{\rm m}$ 03.57$^{\rm s}$, -19$^\circ$ 24\arcmin 51.00\arcsec
(J2000), and has an area of $\approx$ 22 square arcminutes (99.7\%
confidence region).  The prompt localization of this short, hard GRB by
\HETE and the anti-Sun pointing of the \HETE instruments, coupled with
the refinement of the localization by the IPN, has made possible rapid
follow-up observations of the burst at radio, optical, and X-ray wavelengths. 
The time history of GRB020531 at high ($> 30$ keV) energies consists 
of a short, intense spike followed by a much less intense secondary
peak.  Its time history is thus similar to that seen in many short,
hard bursts.  Analysis of the FREGATE and WXM time histories gives
durations for the burst of $t_{50}$ = 1.36 s in the WXM 2 - 25 keV
energy range, and 1.10 s, 0.86 s, 0.62 s, and 0.36 s in the FREGATE
6-13, 14-30, 31-84, and 85-400 keV energy bands.  The duration of the
burst thus increases with decreasing energy, which is similar to the
behavior of long GRBs.  The photon number flux, photon energy flux, and
energy fluence of the burst  in the 50-300 keV energy band in 1.25
seconds are 3.0 ph cm$^{-2}$ s$^{-1}$,  $6.4 \times 10^{-7}$ erg
cm$^{-2}$ s$^{-1}$, and $8.0 \times 10^{-7}$ erg cm$^{-2}$,
respectively.  The spectrum of the burst evolves from hard to soft,
which is also similar to long GRBs.  These similarities to the
properties of long GRBs, and other similarities previously known,
suggest that short, hard GRBs are closely related to long GRBs.
\end{abstract}

\keywords{gamma rays: bursts (GRB020531)}

\section{Introduction}

It has been known for nearly a decade that gamma-ray bursts (GRBs)
appear to fall into two classes: short ($\approx$ 0.2 sec), harder
bursts, which comprise 20-25\% of all bursts; and long ($\approx$ 20
sec), softer bursts, which comprise 75-80\% of the total (Hurley et al.
1992; Lamb, Graziani, and Smith 1993; Kouveliotou et al. 1993). 
Norris, Scargle \& Bonnell 2000).   The spectra of the two classes of
bursts differ: the spectra of the bursts become softer as the bursts
become longer (Dezalay et al. 1992; Kouveliotou et al. 1993; Dezalay et
al. 1996). There is also evidence that the brightness distributions of
the two classes of bursts differ (Graziani and Lamb 1994; Belli 1997;
Tavani 1998); however, the difference in the brightness distributions
can be explained by the difference in their durations, which causes the
sampling distance for short bursts to be smaller than for the long
bursts (Graziani and Lamb 1994).  In addition the $V/V_{\rm max}$
values (Schmidt 2001) and the angular distributions (Kouveliotou et al.
1993) of the two classes appear to be identical.

Thanks to the rapid dissemination of accurate GRB localizations by
\BeppoSAX (Costa et al. 1997), much has been learned in the past five
years about GRBs.  This has included the discoveries that GRBs have
X-ray (Costa et al. 1997), optical (van Paradijs et al. 1997), and
radio  (Frail et al. 1997) afterglows.  Redshifts and host galaxies are
now known for more than two dozen GRBs (see, e.g., Lamb 2002). 
However, all of these discoveries relate to long GRBs.

In contrast, to date nothing is known about the distance to or the
nature of the short GRBs, despite extensive efforts.  The Burst and
Transient Source Experiment (BATSE) on the \CGRO localized localized
numerous short GRBs in near-real time.  Although many had large error
boxes, one (trigger 6788) was localized to an error circle of $\sim 30$
square degrees, which was searched for an optical counterpart within
~12 s to a magnitude of 14.98 (Kehoe et al. 2001).  The results were
negative.  The Third Interplanetary Network (IPN) derived localizations
for four short GRBs (000607, 001025B, 001204, and 010119) with delays
of 15--65 hours.  But in three of these cases, the opportunity for
follow-up observations was compromised either by the burst being close
to the Sun (000607, $65^\circ$) or close to the Galactic plane
(001025B, $b \approx 4^\circ$; 010119, $b \approx 5^\circ$).  Only one
burst (001204) was optimally placed on the sky for follow-up
observations.  However, in this case the delay (65 hours) in deriving a
localization for the burst hampered follow-up efforts.  Despite an
accepted BeppoSAX ToO program, no X-ray  follow-up observations were
possible because of Sun-angle or other operational constraints, except
in the case of 001204.  However, the delay in deriving the localization
of this burst made the success of any X-ray follow-up observation
unlikely, and therefore none was carried out.

In this Letter we report the detection and prompt localization of a
short, hard GRB by {\it HETE-2} (hereafter \HETE) (Ricker et al.
2002a,b).  On 31 May 2002 at 1578.73 SOD (05:15:50.56) UT on 31 May
2002 UT, the HETE-2 French Gamma Telescope (FREGATE) instrument (Atteia
et al. 2002) and the Wide-field X-ray Monitor (WXM) instrument (Kawai
et al. 2002) detected a bright, short (duration $\sim 300$ msec in the
FREGATE 30-400 keV energy band), hard GRB.  The prompt localization of
this short, hard GRB by \HETE and the anti-Sun pointing of the \HETE
instruments has made possible rapid follow-up observations of it
at radio, optical, and X-ray wavelengths.  We also describe the
properties of GRB020531 derived from observations of it using the
FREGATE and WXM instruments on \HETE, which provided unprecedented
spectral and temporal coverage of this short, hard GRB.

\section{Observations} \label{observations}

\subsection{Localization} \label{localization}

The \HETE FREGATE and WXM instruments detected a short (duration $\sim
300$ msec in the FREGATE 30-400 keV energy band), hard gamma-ray burst
(GRB) that occurred at 1578.72 SOD (00:26:18.72 UT) on 31 May 2002. 
The fluence of the burst was not large enough for it to be detected by
the \HETE Soft X-Ray camera (SXC) (Monnelly et al. 2002).  The WXM
flight localization software produced a valid location in spacecraft
(relative) coordinates.  However, since no on-board real-time star
camera aspect was available, an absolute localization could not be
disseminated.  A preliminary ground-analysis localization was reported
as a GCN Position Notice at 01:54:22 UT, 88 minutes after the burst. 
Further ground analysis produced a refined localization, which can be
expressed as a 90\% confidence rectangle that is 67 arcminutes in RA
and 43 arcminutes in Dec, centered at RA = +15$^{\rm h}$ 14$^{\rm m}$
45$^{\rm s}$, Dec = -19$^\circ$ 21\arcmin 35\arcsec (J2000) (see Figure
1).  The refined localizaton was reported as a GCN Position Notice at
03:57:14 UT, 211 minutes after the burst.  

By restricting the search to the crossing window defined by the \HETE
position, a weak GRB was identified in the \Ulysses data.  An initial
IPN localization of the burst was derived by triangulation using \HETE
FREGATE, \Ulysses GRB and \MO HEND data, and a 46 square arcminute
error box was disseminated 18 hours after the GRB (Hurley et al.
2002b).  This localization was further refined approximately five days
after the burst.  The resulting hexagonal-shaped localization error
region is centered on RA = 15$^{\rm h}$ 15$^{\rm m}$ 03.57$^{\rm s}$,
-19$^\circ$ 24\arcmin 51.00\arcsec (J2000), and has an area of
$\approx$ 22 square arcminutes (99.7\% confidence region) (see Figure
1) (Hurley et al. 2002c). 

We have used the imaging capabilities of the WXM to greatly improve the
S/N of GRB020531 as seen by the WXM.  Using the WXM photon time- and
energy-tagged data (TAG data), we selected only WXM photons (1) during
the portion of the burst that maximized the S/N of the burst time
history, using Spiffy Trigger (Graziani 2002); (2) on the seven wires
illuminated by the GRB but not by the bright Galactic X-ray source Sco
X-1, which was in the field-of-view (FOV) of the WXM at the time; (3)
on those portions of these seven wires that were illuminated by the
burst; and (4) the bins on those portions of the wires that were
illuminated by the burst, using the refined IPN localization of the
burst (Hurley et al. 2002b) and the mask pattern of the coded aperture
of the WXM.  These ``cuts'' on the photon TAG data increased the S/N of
the burst in the WXM data by nearly a factor of two.  While these
``cuts'' on the WXM photon TAG data do not allow an improved
localization of the GRB by the WXM, their success provides independent
evidence that the refined IPN error region for the burst is correct.

\subsection{Temporal Properties} \label{time_history}

Figure 2 shows the time history of GRB020531 in the WXM 2-25 keV energy
band and in various FREGATE energy bands.  Table lists the $t_{50}$ and
$t_{90}$ durations of the burst in the WXM 2-25 kev energy band and in 
various FREGATE energy bands.

Figure 3 shows the burst time history in the entire WXM 2-25 keV energy
band, and in four other WXM energy bands, utilizing the four ``cuts''
on the WXM photon TAG data that increase the S/N of the burst in the
WXM data, and that are described in \S2.1 section.

The $t_{50}$ and $t_{90}$ durations of GRB020531 in the 85-400 keV
energy band were 0.36 s and 0.74 s; thus GRB020531 is a short, hard
GRB.  Like many short bursts (see, e.g., 010119; Hurley et al. 2002),
the time history of GRB020531 consists of a short, intense spike
followed by a less intense and softer secondary peak.  At low energies,
the burst consists of two peaks, each lasting $<$ 1 sec and separated
by about 2 seconds (see Figure 3).  Thus, at low energies, the primary
and secondary peaks are comparable in duration and in intensity.

Figure 4 shows the $t_{50}$ and $t_{90}$ durations of GRB020531 in 
various energy bands.  The duration of the burst increases with
decreasing energy.  A $\chi^2$ fit to the $t_{50}$ values, assuming
that all of the values have the same relative uncertainty, yields $\log
t_{50} = 0.42 -  0.38 \log (E/{1 {\rm keV}})$; a similar fit to the
$t_{90}$ values yields $\log t_{90} = 1.42 - 0.62 \log (E/{1 {\rm
keV}})$. This behavior is similar to that seen in long GRBs (Fenimore
et al. 1994).

\subsection{Spectrum} \label{spectrum}

Table 3 gives the S/N of the detection of the burst, the peak photon
flux, the peak energy flux, and the energy fluence of the burst in
the WXM and in various FREGATE energy band.

Table 4 gives the best-fit parameters for the spectrum of GRB020531 in
the WXM and in FREGATE.  Figure 5 shows the expected and observed
photon counts in FREGATE energy loss bins from 6--400 keV (upper panel)
and the residuals (lower panel) for the best fit power-law spectrum
given in Table 4.  Figure 6 shows the corresponding photon counts in
the WXM energy loss bins from 2-25 keV for the four wires in the
X-detector and the two wires in the Y-detector (Kawai et al. 2002)
used in the WXM spectral fit.

The spectra derived independently in the WXM and in FREGATE are
consistent with one another.  The results show that the observed
spectrum of GRB020531 is fully consistent with a single power-law
spectrum with slope $\alpha = -1.2 \pm 0.06$.  The power-law spectrum
of GRB020531 extends from 2--400 keV, and shows no evidence of a cutoff
at high energies to the highest energies ($\approx$ 400 keV) observed
by FREGATE.

We have explored the spectral evolution of the burst in the WXM 2--25
keV energy band.  Using Spiffy Trigger (Graziani 2002), we found two
time intervals during the burst that maximized the S/N of the time
history.  These two time intervals start 0.00 sec and 1.12 sec after
the trigger time for the burst, and last $\Delta t_1 = 0.80$ sec and 
$\Delta t_2 = 0.72$ sec.  We have used the four ``cuts'' on the WXM
photon TAG data that are described in \S2.1, and that increase the S/N
of the burst in the WXM data, in order to determine the spectrum of the
burst in the WXM 2--25 keV energy band in these two time intervals. 
Table 4 shows the resulting best-fit power-law parameters for these two
time intervals.  Comparing the value of the  power-law index ($\alpha =
1.26 \pm 0.06$) for the spectrum during the first peak, as determined
from the FREGATE data, and the value of the  power-law index ($\alpha =
2.186^{+1.38}_{-0.95}$) for the spectrum during the second peak, as
determined from the WXM data, provides evidence at the $\approx$ 90\%
confidence level that the spectrum of the secondary peak is softer than
the spectrum of the primary peak.  Such spectral softening is similar
to that seen in long GRBs (see, e.g., Band et al. 1993).

\section{Discussion} \label{discussion}

\HETE has detected and localized GRB020531, a short, hard burst.  The
prompt localization of the burst by \HETE and the anti-Sun pointing of
the \HETE instruments, coupled with the later precise localization of
the burst by the IPN, has allowed rapid follow-up of the GRB not only
by small aperture, large FOV robotic telescopes (e.g., Park et al.
2002, Boer et al. 2002) but also by large aperture, modest FOV
telescopes (e.g., Fox et al. 2002, Lamb et al. 2002, West et al. 2002,
Miceli et al. 2002, Dullighan et al. 2002).  Figure 7 shows that these
observations have placed much more severe upper limits on any optical
afterglow of a short, hard GRB than ever before.  However, these
constraints do not rule out the existence of optical afterglows of
short, hard GRBs that are similar to the optical afterglows of long
GRBs (see Hurley et al. 2002, Figure 3).

The time history of GRB020531 at high ($> 30$ keV) energies consists 
of a short, intense spike followed by a much less intense secondary
peak.  Its time history is thus similar to that seen in many short,
hard bursts.  The time history of GRB020531 at low ($< 25$ keV)
energies consists  of two peaks of similar duration ($\approx$ 0.80
sec) and intensity, the first of which corresponds in time to the
short, intense spike seen at high energies.

The spectrum of the short, intense spike is well described by a single
power-law spectrum with index $\alpha = - 1.2 \pm 0.06$ from 6--400
keV.  Such a steep spectrum is quite unusual (Paciesas et al. 2001) but
not unprecedented (see, e.g., Hurley et al. 2002; Figure 2 [GRB001204])
for short, hard GRBs.  The steepness of the spectrum of GRB020531 may
explain in part the fact that the WXM on \HETE was able to detect and
localize this particular short, hard burst, whereas the Wide-Field
Cameras on \BeppoSAX were ultimately unsuccessful in dectecting or
localizing any short, hard GRBs, despite great efforts (Gandolfi et al.
2002).

The spectrum of the secondary peak, which is comparable in duration and
in intensity in the 2--25 keV energy band to the primary peak, is also
well described by a single power-law spectrum, but the spectrum is
softer than the spectrum of the primary peak at the 90\% confidence
level.

Two qualitatively different models have been suggested to explain
GRBs: merging compact objects and the collapse of rotating massive
stars. It is difficult to make short bursts in the collapsar model
(MacFadyen \& Woosley 1999) and there have been suggestions (e.g.,
Fryer, Woosley, \& Hartmann 1999) that short hard bursts are the
observational counterpart of merging neutron-star and neutron-star
black-hole pairs. While collapsars and merging compact objects both
get their energy from a hyperaccreting black hole, the dimension and
mass of the disk is smaller in the latter, hence the time scale is
shorter. If these mergers go on far from the galaxies where the
compact objects are made, their afterglows might be faint.

The spectral and temporal behavior measured by \HETE for GRB020531
could indicate a central engine that remains on with a declining power
after the principal burst (reflecting a declining accretion rate?) or
may be a consequence of the competition between internal and external
shocks in making the GRB. The short time scale is certainly more
consistent with the merging compact object hypothesis, but also
possibly consistent with the supranova model (Vietri \& Stella 1998,
1999) since the latter produces a similar compact disk and black hole. 
This could give a short burst should such bursts prove to be associated
with massive stars. 

The properties of GRB020531 as measured by \HETE have different 
characteristics at different energies: more complex time structure at
lower energies, increasing duration with decreasing energy, a power-law
spectrum over the 2-400 keV energy range but spectral softening with 
time).  These properties of GRB020531 as measured by \HETE are similar
to those of long bursts, and when  taken together with the previously
know properties of short, hard GRBs described in the Introduction
(similar brightness distributions, $V/V_{\rm max}$ values, angular
distributions) suggest that short, hard GRBs are closely related to
long GRBs.

\section{Conclusions} \label{conclusions}

\HETE has detected and localized a short, hard GRB, establishing its
capability to do so -- and demonstrating that the detection and
localization of short, hard GRBs in the hard x-ray energy band is
possible.  This has important implications for Swift and for other
future GRB missions.

The prompt, precise localization of GRB020531 by \HETE and the IPN have
allowed rapid follow-up observations, which have placed much more
severe limits on the brightness of any radio and optical afterglows
from short, hard GRBs.

The complement of soft x-ray, hard x-ray, and gamma-ray instruments
(SXC, WXM, and FREGATE) on \HETE provides unprecedented  temporal and
spectral coverage of short, hard GRBs.  With the currently projected
long orbital lifetime (> 10 yrs) and excellent health of the \HETE
spacecraft and instruments, the results described for GRB020531 in this
Letter demonstrate that \HETE can continue to provide an unprecedented
opportunity to study short, hard GRBs, and possibly to determine the
distance to and the nature of these bursts.

\acknowledgments
\section*{Acknowledgments}

The HETE mission is supported in the US by NASA contract NASW-4690; in
Japan, in part by the Ministry of Education, Culture, Sports, Science,
and Technology Grant-in-Aid 13440063; and in France, by CNES contract
793-01-8479.  KH is grateful for \Ulysses support under Contract JPL
958056, for \HETE support under Contract MIT-SC-R-293291, and for \MO
support under the NASA LTSA program.  G. Pizzichini acknowledges
support by the Italian Space Agency.

\clearpage

\begin{deluxetable}{lccc}
\tablecaption{GRB020531 Error Box Coordinates}
\tablewidth{0pt}
\tablehead{
\colhead{Source} & \colhead{$\alpha_{\rm J2000.0}$} & 
\colhead{$\delta_{\rm J2000.0}$} & \colhead{Comment} 
}
\startdata
HETE WXM & 15 14 45 &      -19 21 35 & center \\
          & 15 17 00 &      -19 43 00 & corner \\
          & 15 17 00 &      -19 00 00 & corner \\
          & 15 12 30 &      -19 00 00 & corner \\
          & 15 12 30 &      -19 43 00 & corner \\
&&& \\
IPN & 15 15 03.57 &   -19 24 51.00 & center \\
    & 15 14 53.98 &   -19 24 18.15 & corner \\
    & 15 15 14.46 &   -19 21 38.39 & corner \\
    & 15 15 17.07 &   -19 21 35.32 & corner \\
    & 15 15 12.51 &   -19 25 32.46 & corner \\
    & 15 14 53.75 &   -19 27 58.57 & corner \\
    & 15 14 49.67 &   -19 28 03.27 & corner \\

\enddata
\vskip -18pt
\tablecomments{Units of right ascension are hours, minutes, and
seconds; units of declination are degrees, arcminutes, and arcseconds.}
\end{deluxetable}

\begin{deluxetable}{lccc}
\tablecaption{Temporal Properties of GRB020531.\label{tbl-3}}
\tablewidth{0pt}
\tablehead{
\colhead{Instrument} & \colhead{Energy Band} & \colhead{$t_{50}$} & \colhead{$t_{90}$} \\
& \colhead{(keV)} & \colhead{(s)} & \colhead{(s)}
}
\startdata
HETE WXM & 2 - 25 & 1.36 & 4.56 \\
&&& \\
HETE FREGATE &  6-13 & 1.10 & 5.58 \\
             & 14-30 & 0.86 & 3.46 \\
             & 31-84 & 0.62 & 4.52 \\
             & 85-400 & 0.36 & 0.74 \\
\enddata

\end{deluxetable}

\begin{deluxetable}{rcccc}
\tablecaption{Energy Emission Properties of GRB020531.\label{tbl-1}}
\tablewidth{0pt}
\tablehead{
\colhead{Energy Band} & \colhead{Photon Flux} & \colhead{Energy Flux} & 
\colhead{Energy Fluence} \\
\colhead{(keV)} & \colhead{(ph cm$^{-2}$ s$^{-1}$)} & 
\colhead{(erg cm$^{-2}$ s$^{-1}$)} & \colhead{(erg cm$^{-2}$)}
}
\startdata
  2 - 25 &  2.9 &  $(2.7 \pm 0.8) \times 10^{-8}$ &
$(4.9^{+1.5}_{-1.4}) \times 10^{-8}$ \\
 8 - 400 &  8.4 &  $10.1 \times 10^{-6}$ & $12.7 \times 10^{-7}$ \\
32 - 400 & 4.4 &  $9.1 \times 10^{-7}$ & $11.4 \times 10^{-7}$ \\
25 - 100 & 3.0 &  $2.6 \times 10^{-7}$ & $3.1 \times 10^{-7}$ \\
50 - 300 & 3.0 &  $6.4 \times 10^{-7}$ & $8.0 \times 10^{-7}$ \\

\enddata
\vskip -18pt
\tablecomments{All WXM parameters are for a joint fit to six wires (XA0,
XA1, XA2, XB0, YB0, and YB1).}
\end{deluxetable}

\begin{deluxetable}{rcccc}
\tablecaption{Parameters for Best-Fit Power-Law Spectrum of GRB020531.\label{tbl-1}}
\tablewidth{0pt}
\tablehead{
\colhead{Instrument} & \colhead{$\Delta t$} & \colhead{Energy Band} & 
\colhead{Scale Factor} & \colhead{Power-Law Index} \\
& \colhead{(s)} & \colhead{(keV)} &  & \colhead{$\alpha$} 
}
\startdata
WXM & 0.00-1.84 & 2 - 25 & $3.936^{+0.99}_{-0.93}$ & $1.57^{+0.48}_{-0.44}$ \\
& 0.00-0.80 & " & --- & $1.15 ^{+0.80}_{-0.54}$  \\
& 1.12-1.84 & " & --- & $3.06^{+2.55}_{-1.27}$ \\
FREGATE & (-0.145)-(+1.31) & 6-400 & ${(8.05 \pm 0.56) \times 10^{-2}}$ & $1.26 \pm 0.06$ \\
\enddata
\vskip -18pt
\tablecomments{The scale factor for the WXM fit is normalized at 1 keV, 
while that for the FREGATE fit is normalized at 30 keV.  The WXM parameters are
for a joint fit to six wires (XA0, XA1, XA2, XB0, YB0, and YB1).}

\end{deluxetable}

\begin{figure} \plotone{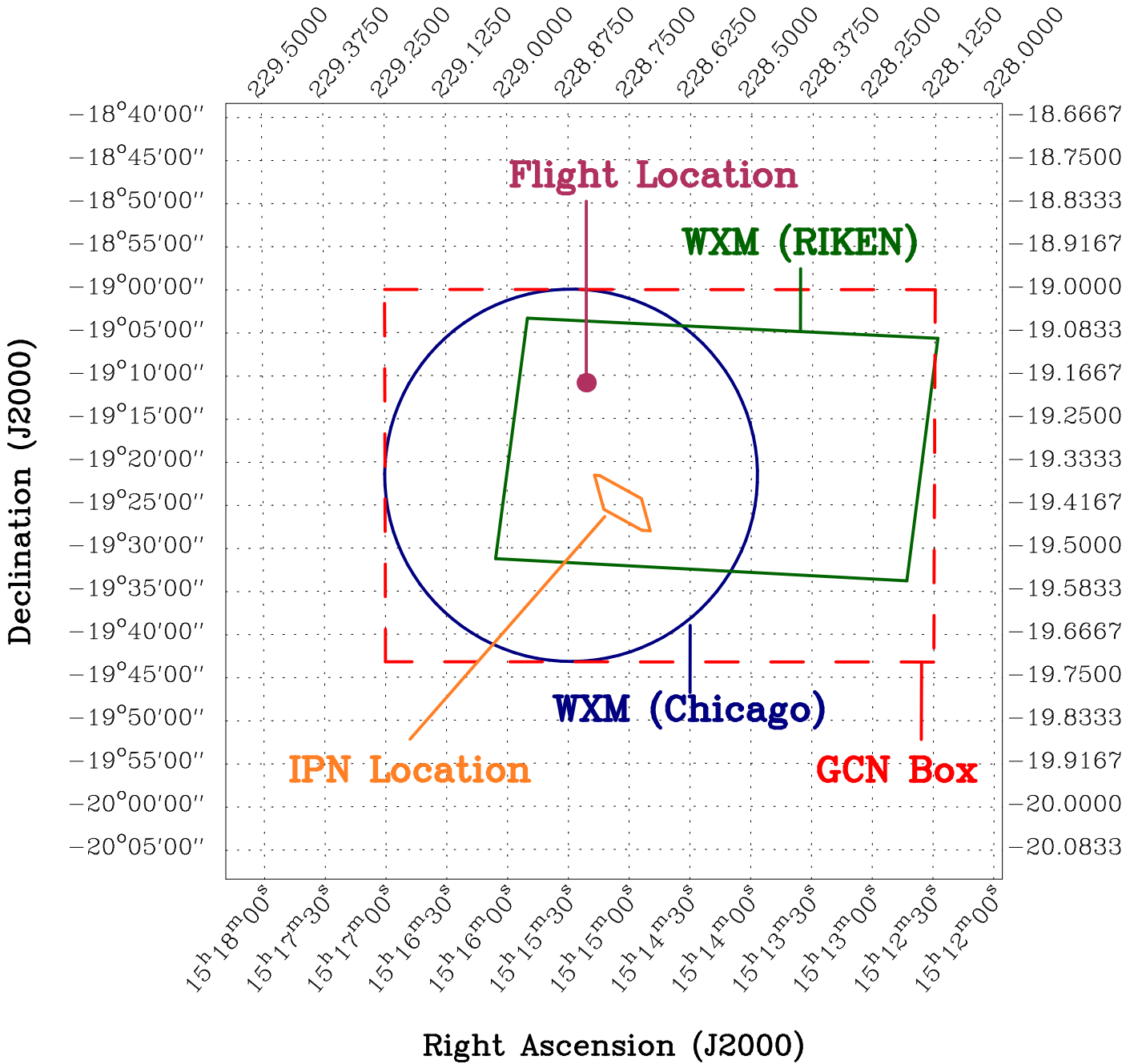} \caption{The final
rectangular HETE WXM error box for GRB020531 (dashed line).  The
rectangle completely encloses the Chicago 90\% confidence region error
circle and RIKEN 90\% confidence region error rectangle (which utilyzed
the same data).  Note that the WXM flight location lies well inside the
confidence region.  Also shown is the hexagonal-shaped refined IPN
error box for GRB020531 (thin solid line), determined by triangulation
using the \HETE FREGATE, \Ulysses GRB, and \MO HEND data for the burst.
\label{Fig. location}}
\end{figure}

\begin{figure}
\includegraphics
[angle=270,scale=0.3,clip,bb=90 30 576 711]{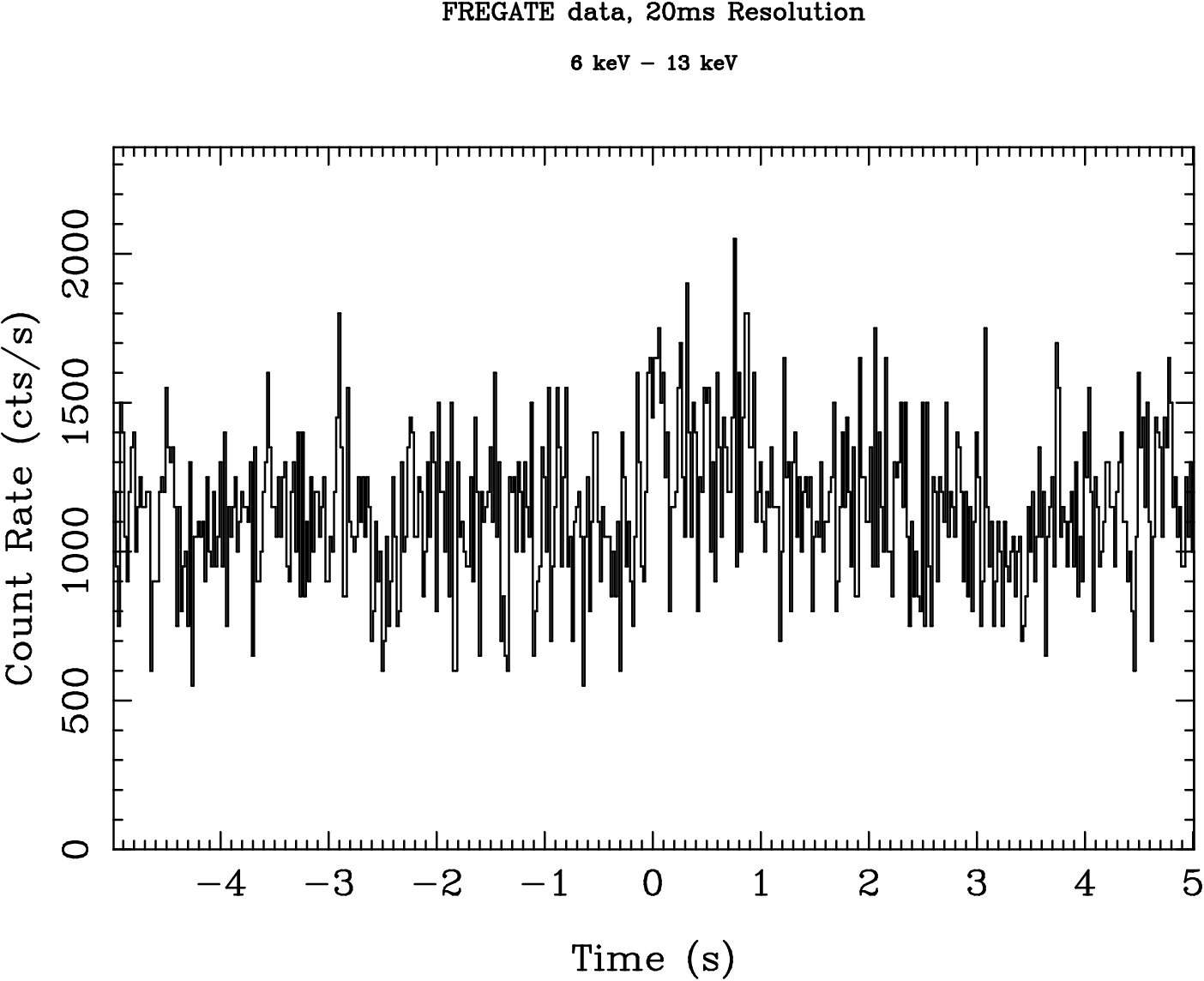}\\
\includegraphics
[angle=270,scale=0.3,clip,bb=90 30 576 711]{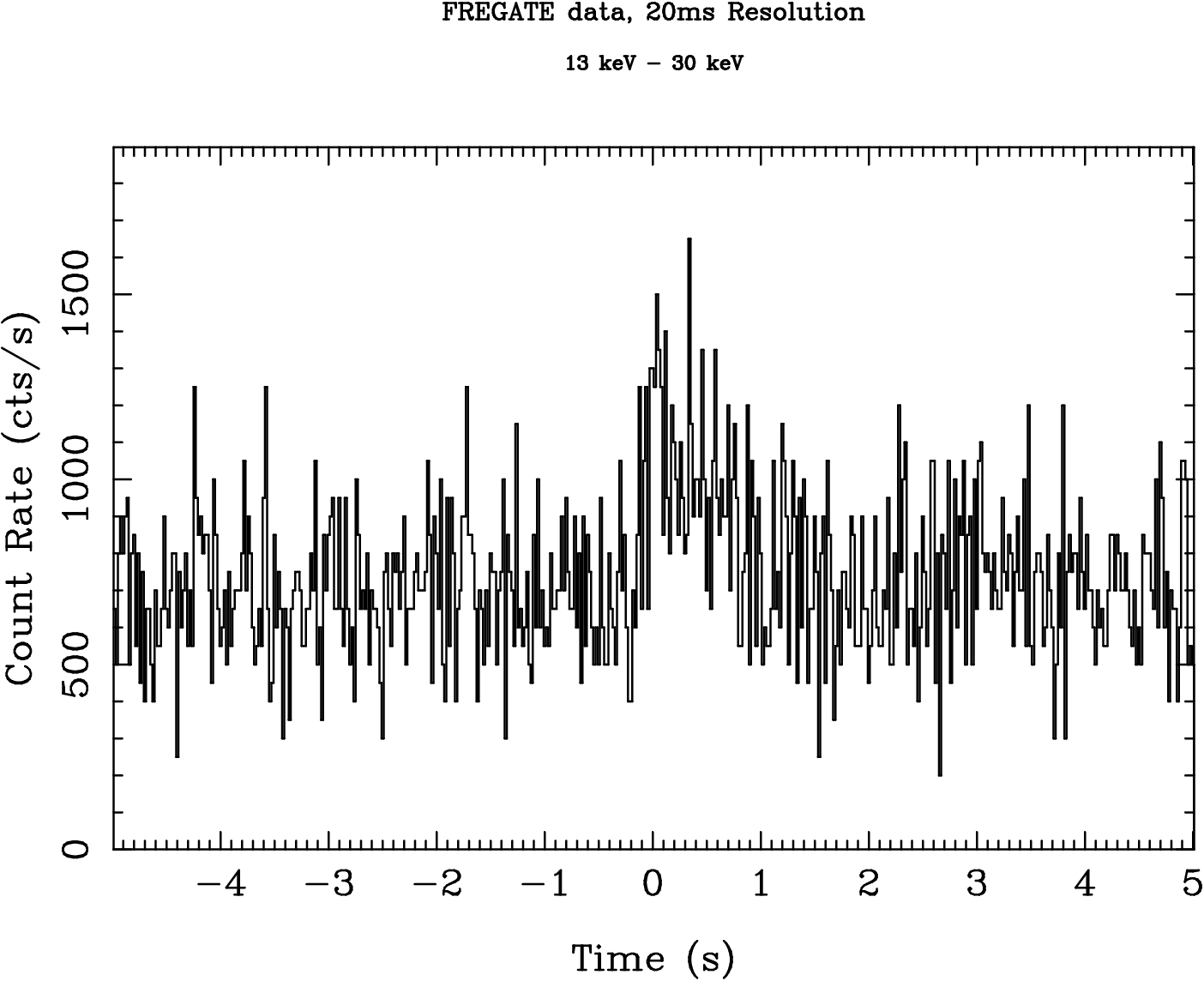}\\
\includegraphics
[angle=270,scale=0.3,clip,bb=90 30 576 711]{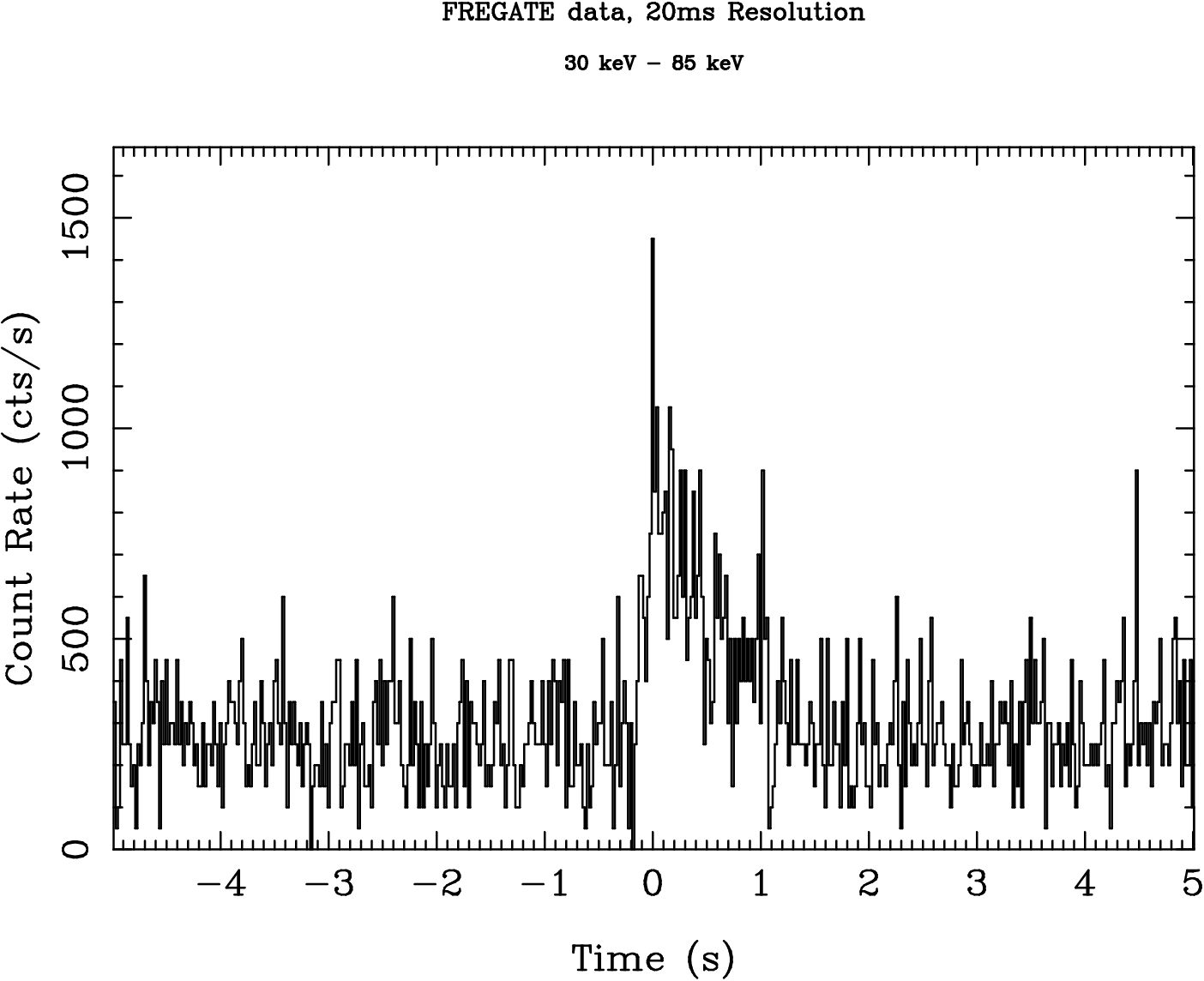}\\
\includegraphics
[angle=270,scale=0.3,clip,bb=90 30 576 711]{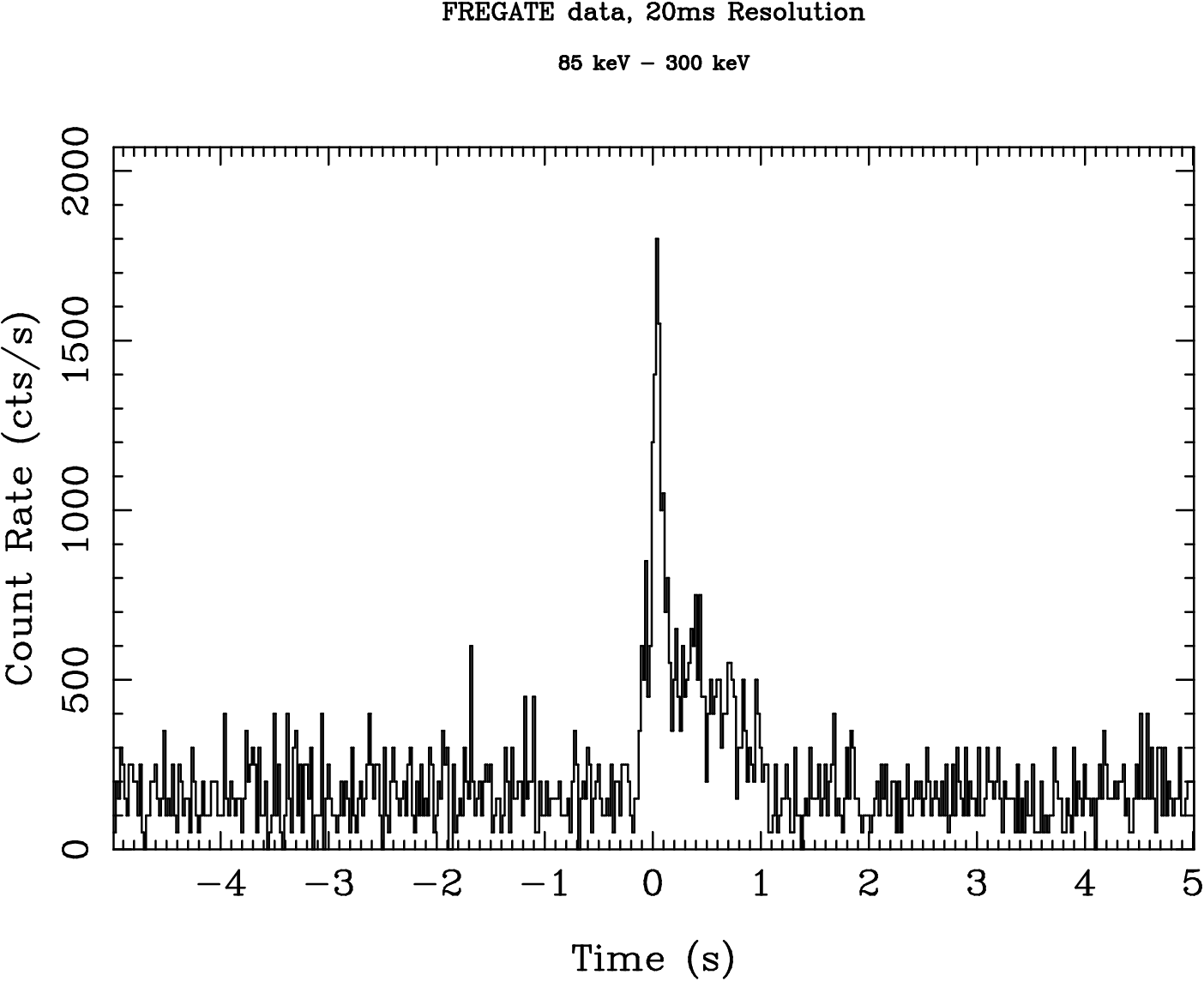}
\caption{FREGATE time histories of GRB020531, binned in 80 msec bins. 
Top to bottom: the 6-13~keV, 13-30~keV, 30-85~keV, and 85-300~keV
bands.}
\label{Fig. light_curves}
\end{figure}

\begin{figure}
\includegraphics
[angle=270,scale=0.3,clip,bb=90 30 576 711]{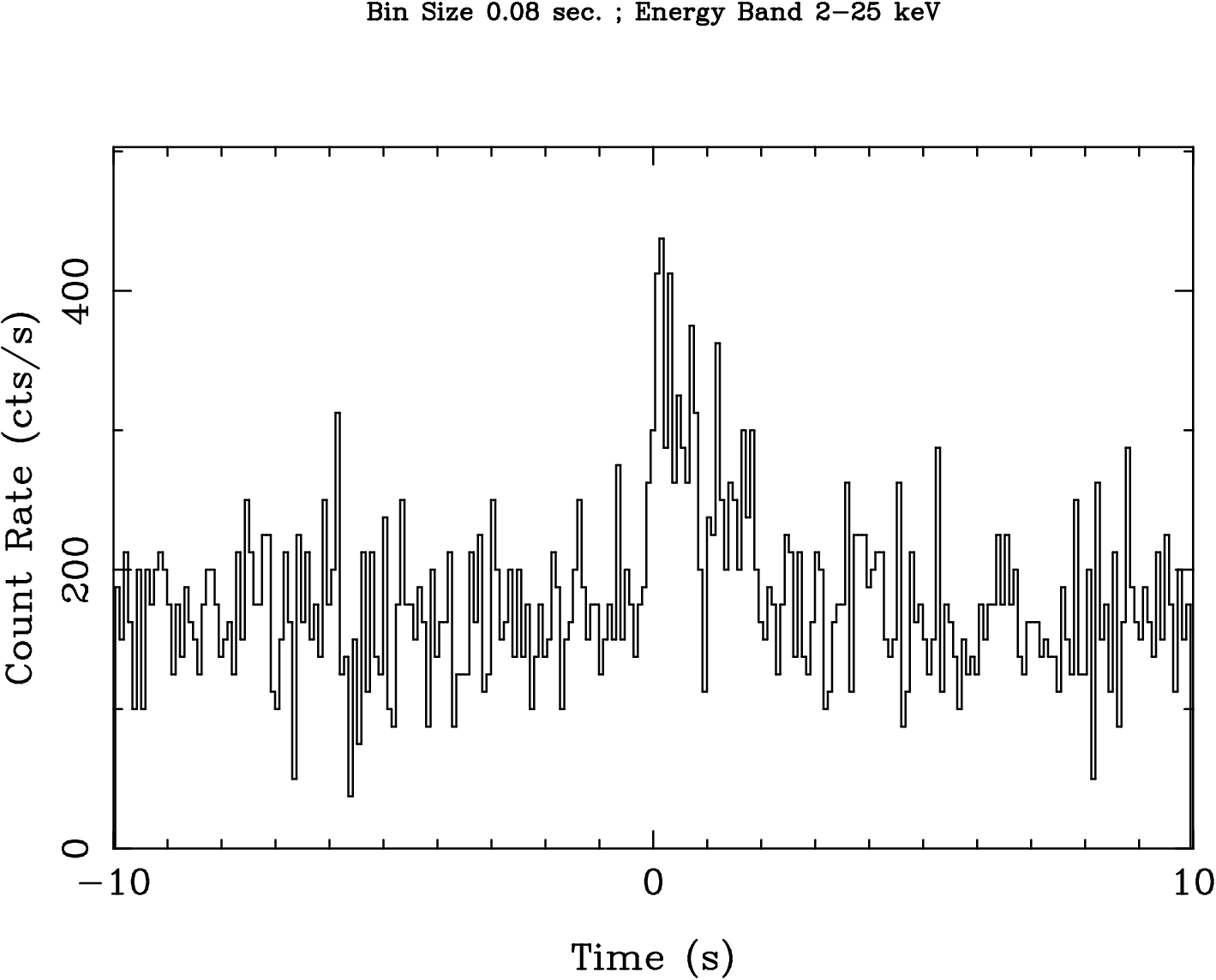}\\
\includegraphics
[angle=270,scale=0.3,clip,bb=90 30 576 711]{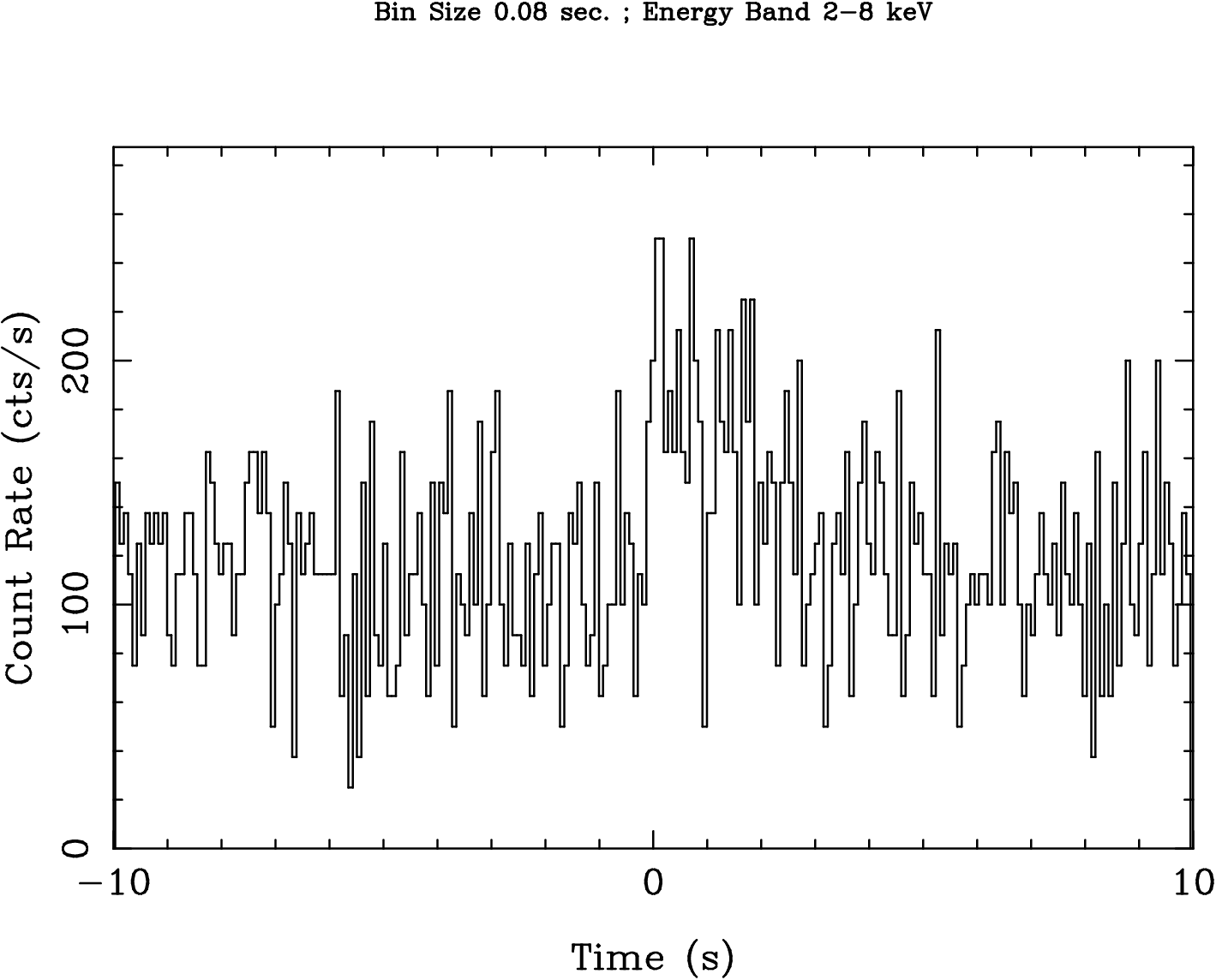}\\
\includegraphics
[angle=270,scale=0.3,clip,bb=90 30 576 711]{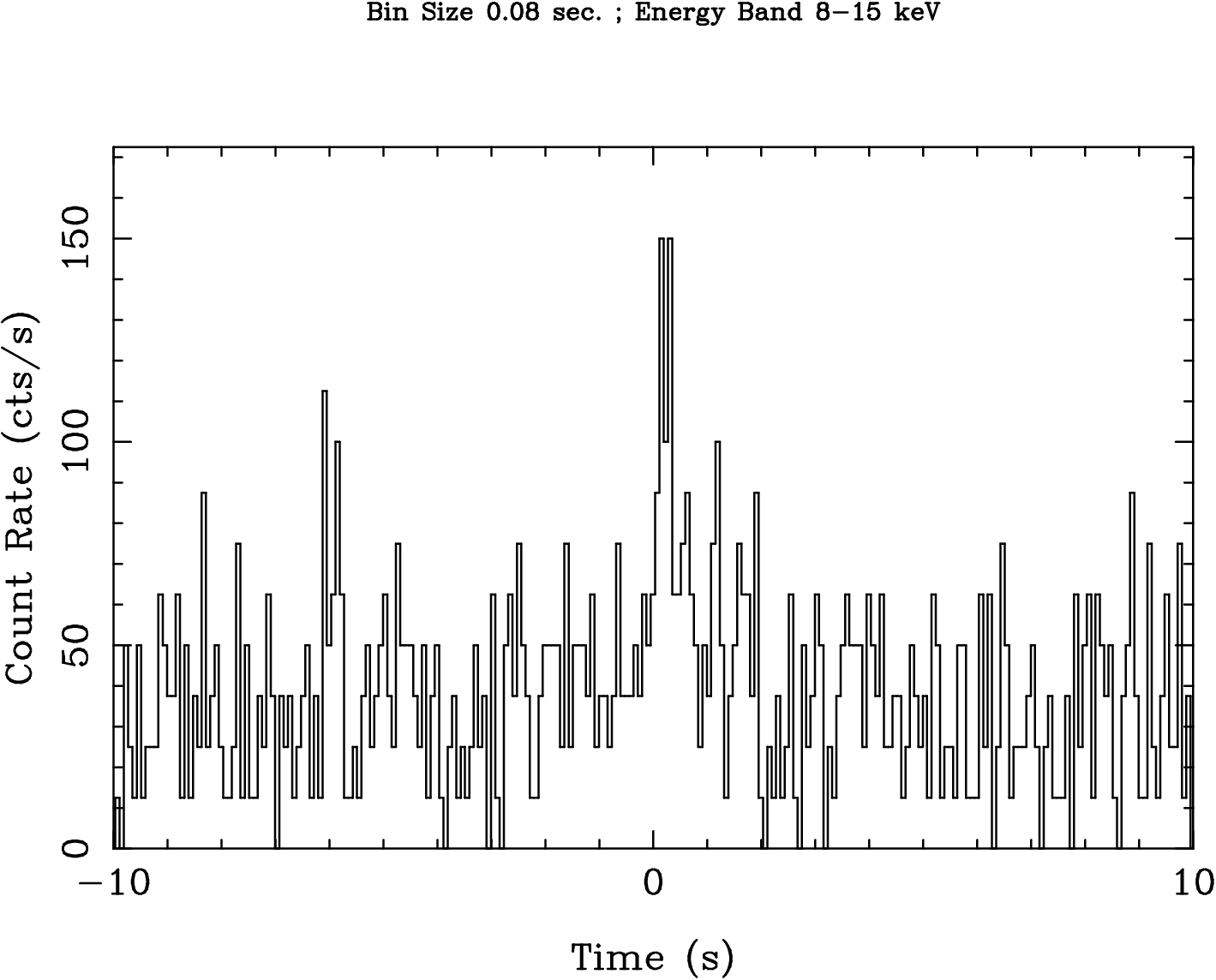}\\
\includegraphics
[angle=270,scale=0.3,clip,bb=90 30 576 711]{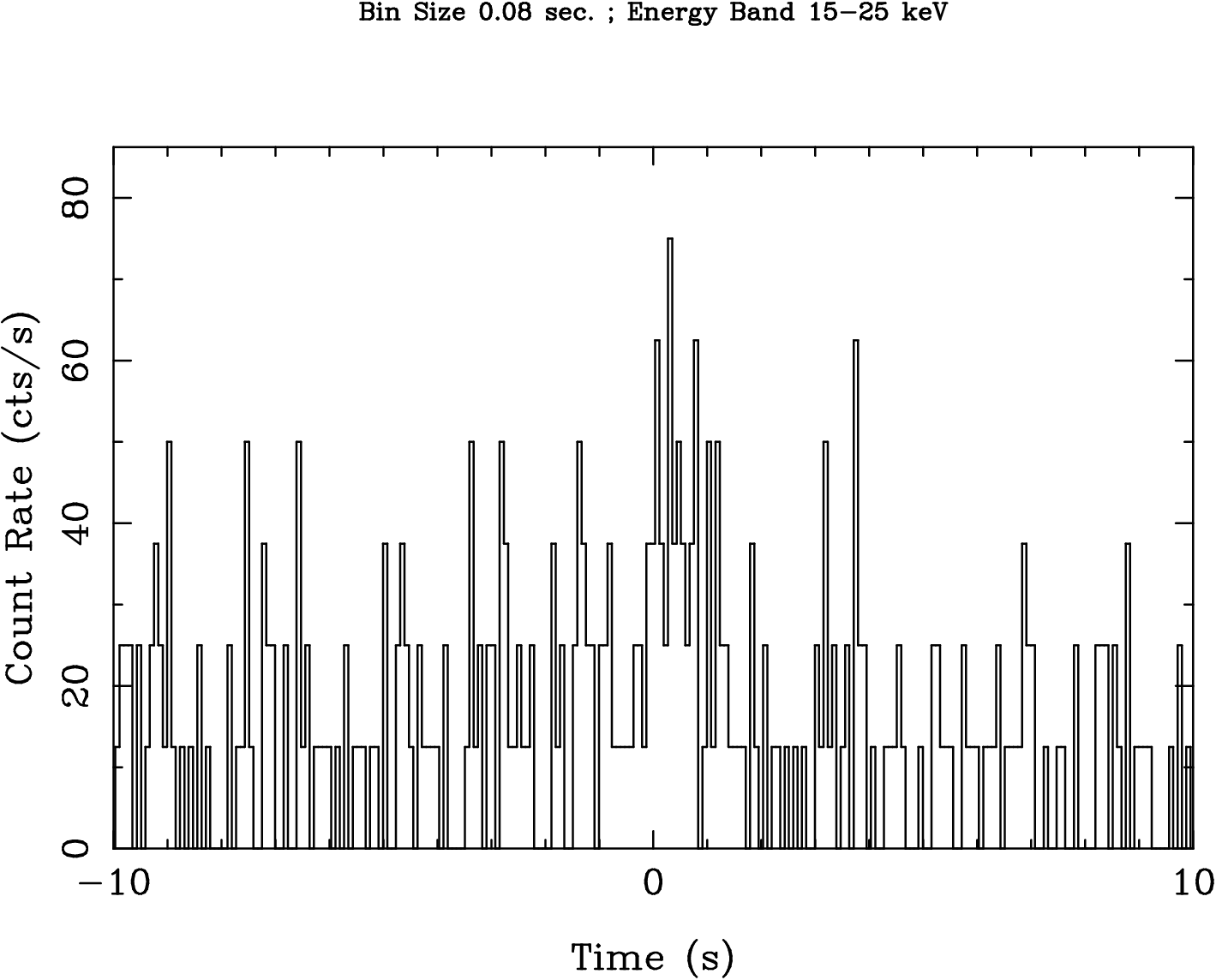}
\caption{Time history of GRB020531 in the WXM 2-25 keV, 2-8 keV, 8-15
keV, and 15-25 keV energy bands, binned in 80 msec bins. \label{Fig.
wxm_light_curves}}
\end{figure}

\begin{figure}
\plotone{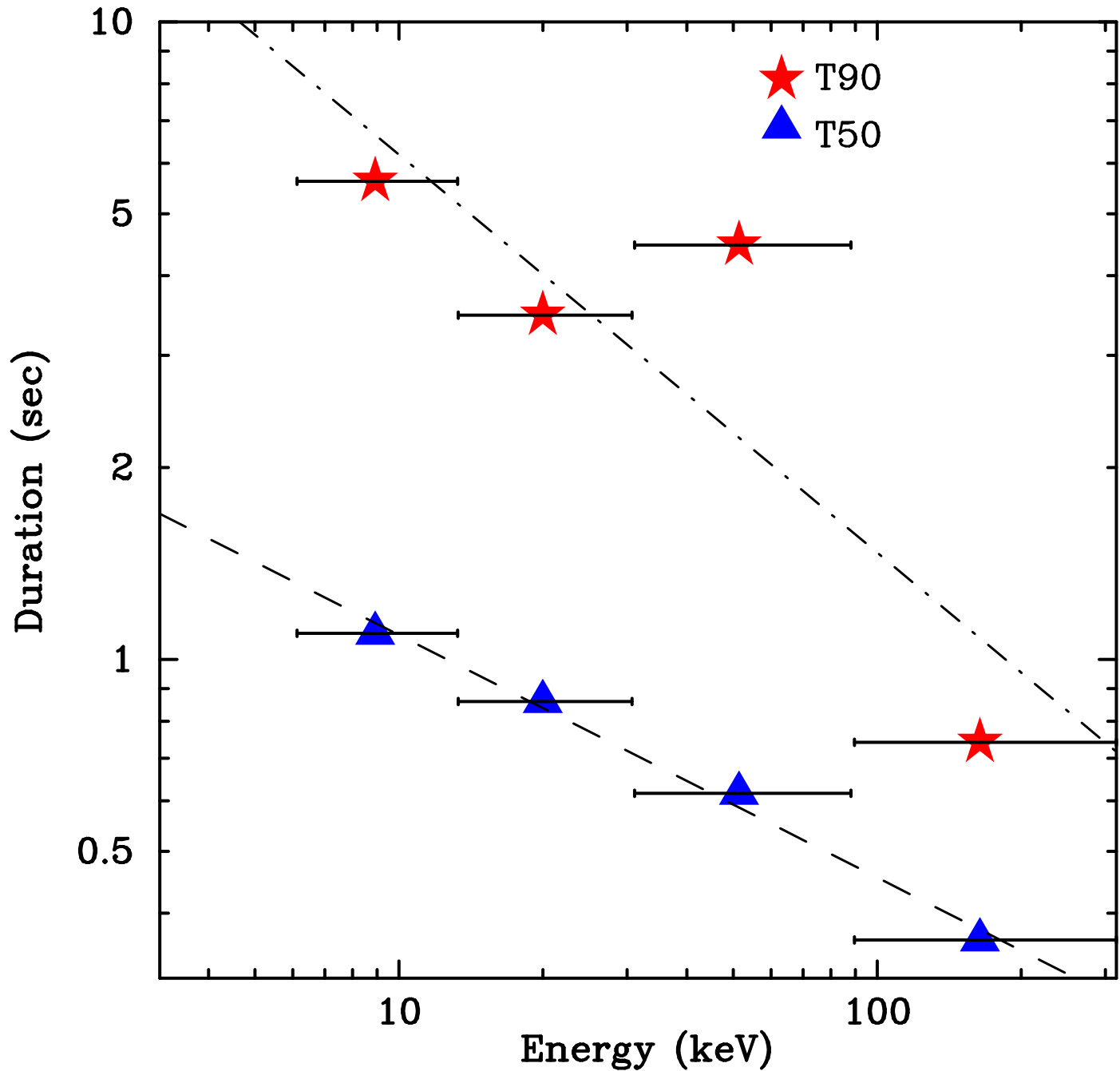}
\caption{Duration of GRB020531 versus energy.  The energy bins have been
chosen so that each contains approximately the same number of photons.  
The durations $t_{50}$ and $t_{90}$ increase with decreasing energy as
$E^{-0.38}$ and $E^{-0.62}$, respectively.  This behavior is similar to
that seen in long GRBs. \label{Fig. e_vs_t}}
\end{figure}

\begin{figure}
\includegraphics[angle=270,scale=0.8]{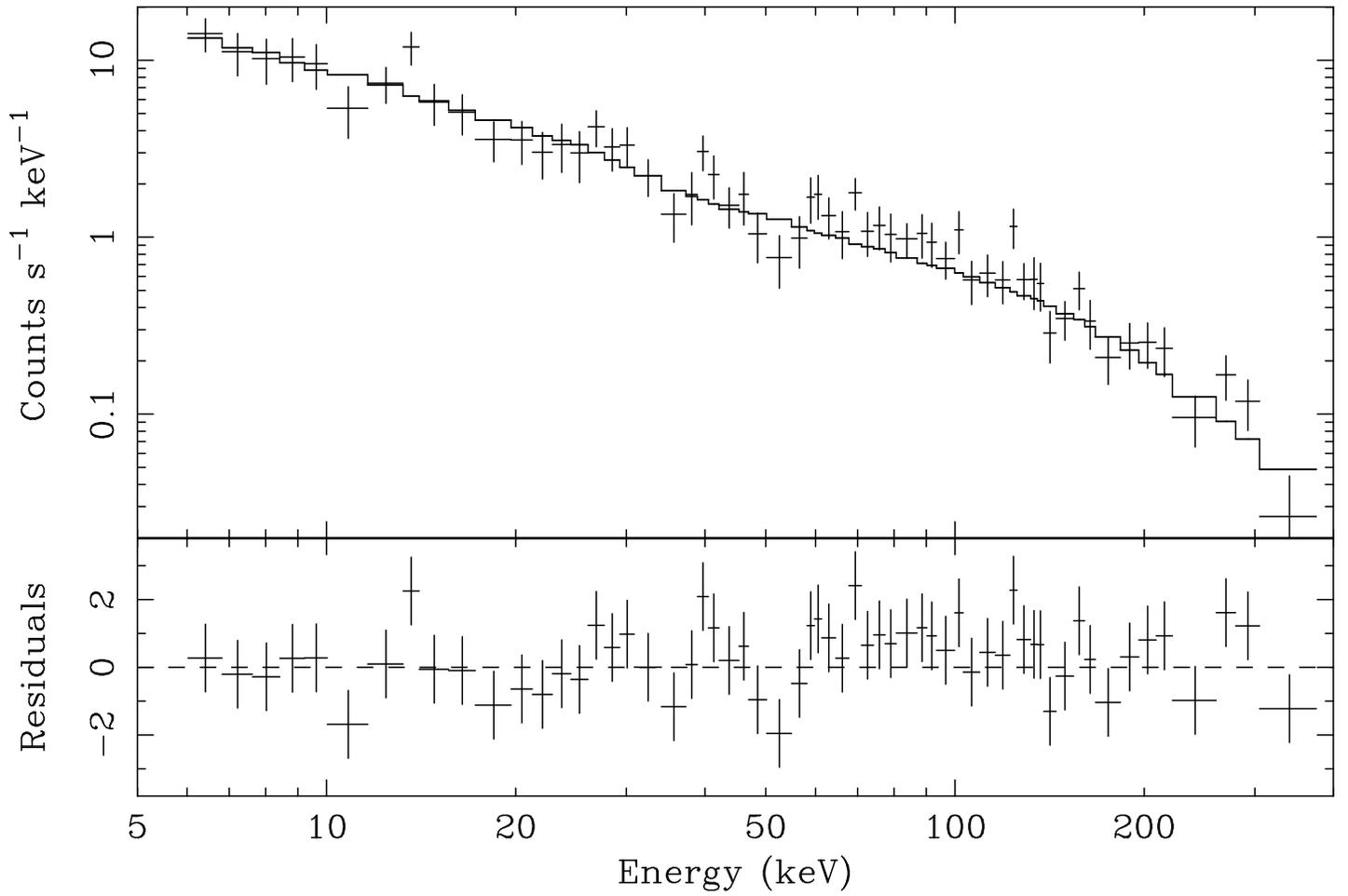}
\caption{Upper panels:  Observed counts (crosses) compared to predicted
counts (histogram) for the best-fit spectrum in the FREGATE energy band
6-400 keV during the first 0.00-1.25 sec of the burst.  Lower panels:
residuals. \label{Fig. fregate_spectrum}}
\end{figure}

\begin{figure}
\includegraphics[angle=270,scale=0.7,bb=110 44 560 710,clip]{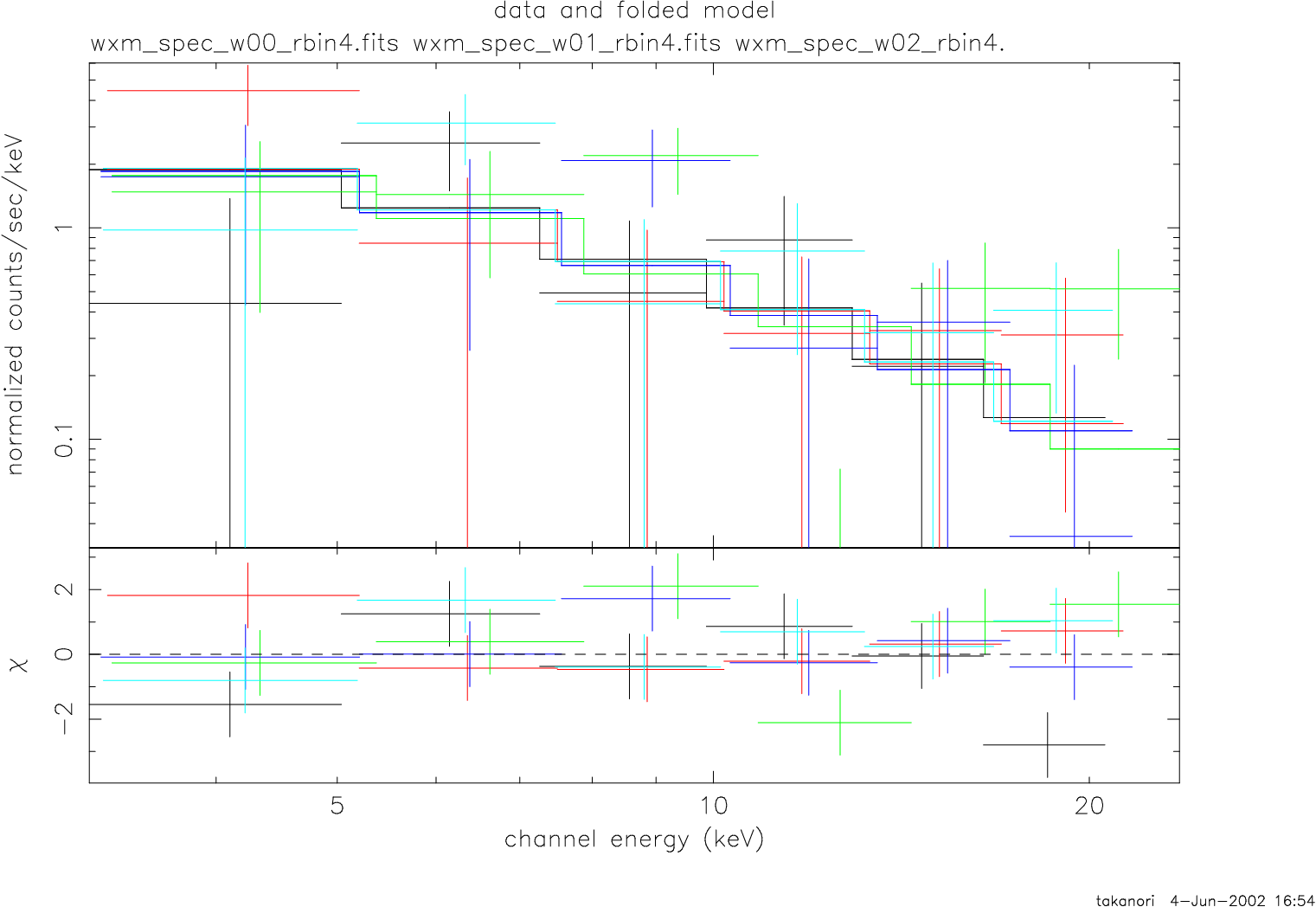}\\
\includegraphics[angle=270,scale=0.7,bb=110 44 560 710,clip]{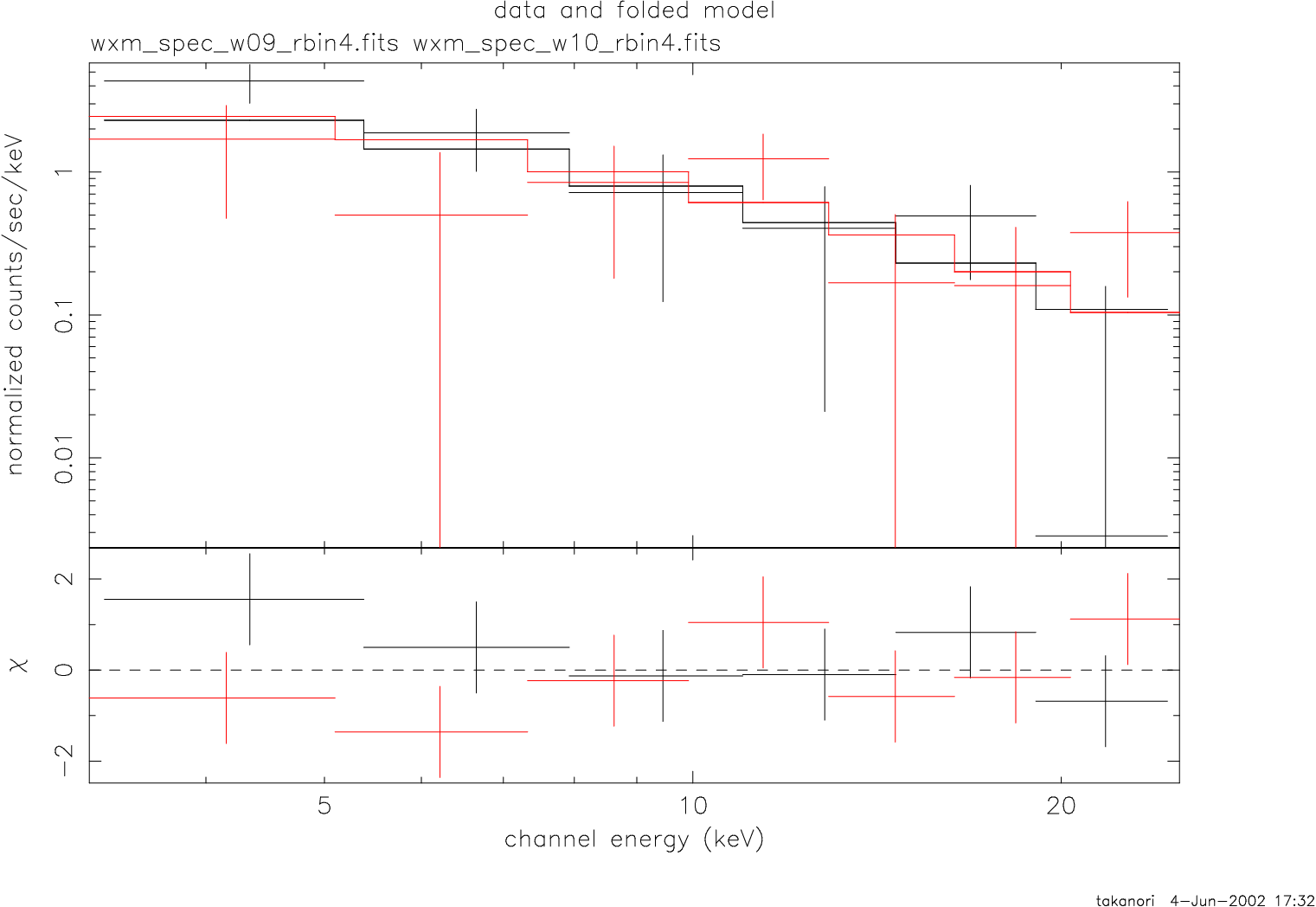}
\hfil
\caption{Observed counts compared to predicted counts for the best-fit
spectrum in the WXM energy band 2-25 keV for the first 1.84 seconds of
the burst, and the residuals.  A specific detector response matrix is
calculated for each anode wire.  Top panels:  Observed counts (crosses)
compared to predicted counts (histograms) for wires XA0, XA1, XA2, and
XB0 (upper panel); residuals (lower panel).  Bottom panels: Observed
counts (crosses) compared to predicted counts (historams) for wires
YB0, and YB1 (upper panel); residuals (lower panel). \label{Fig.
wxm_spectrum}}
\end{figure}

\begin{figure}
\includegraphics[angle=270,scale=0.7]{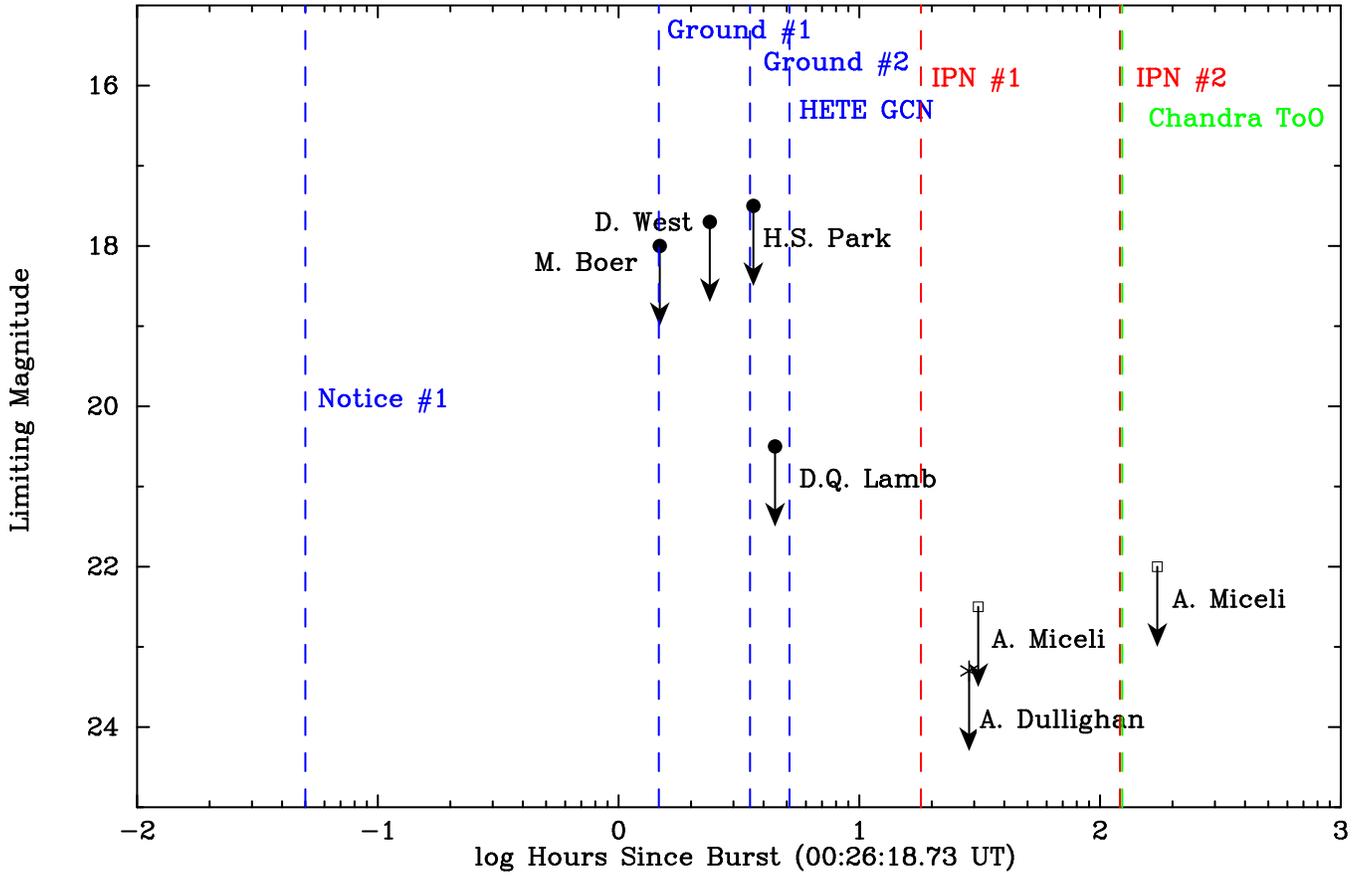}
\caption{Limiting magnitudes versus time for any optical afterglow
from  follow-up observations of GRB020531.  The limiting magnitudes
shown include both limits derived using small aperture, large FOV
robotic telescopes (Park et al. 2002, Boer et al. 2002, West), a small 
aperture, modest FOV telescope (West et al. 2002), and large aperture,
small  FOV telescopes (Lamb et al. 2002, West et al. 2002, Miceli et
al. 2002, Dullighan et al. 2002).  \label{Fig. followups}}
\end{figure}


\begin{thebibliography}{}

\bibitem[Atteia(2002)]{WH2001_FREGATE}
Atteia, J-L, et al. 2002, {\it In-flight Performance and First  Results
from the FREGATE Instrument on HETE}, in WH2001\footnote[7]{WH2001 =
{\it Gamma-Ray  Burst and Afterglow Astronomy 2001: A Workshop
Celebrating the First Year of the HETE  Mission}, Woods Hole, MA,
November 2001, to be published in the AIP Conference  Proceedings (AIP
Press: New York).}.

bibitem[Band1993]{}
Band, D. et al. 1993, ApJ, 413, 281

\bibitem[Belli97)]{}
Belli, B. 1997, in Proc. 25th Int. Cosmic-Ray Conf. (Durban), 41

\bibitem[Boer2002]{}
Boer, M. Klotz, A., Atteia, J.-L., Pollas, C. \& Pinna, H.  2002, GCN
Circular 1408

\bibitem[Butler2002]{}
Butler, N., Dullighan, A., Ford, P., Monnelly, G., Ricker, G.,
Vanderspek, R., Hurley, K. \& Lamb, D.  2002, GCN Circular 1415

\bibitem[Costa1997]{}
Costa, E. et al. 1997, IAU Ciruclar No. 6576

\bibitem[Dezalay92)]{}
Dezalay, J.-P. et al. 1992, in AIP Conf. Proc. 265, Gamma-Ray Bursts,
ed. W. Paciesas \& G. Fishman (New York: AIP), 304

\bibitem[Dezalay96)]{}
Dezalay, J.-P. et al. 1996, ApJ, 471, L27

\bibitem[Dullighan]{}
Dulligan, Monnelly, G., Butler, N., Vanderspek, R., Ford, P. \& Ricker,
G. 2002, GCN Circular 1411

bibitem[Fenimore1994]{}
Fenimore, E. E. 1994, ApJ, 547, 315

\bibitem[Fox2002]{}
Fox, D. \& Bloom, J. S.  2002, GCN Circular 1400

\bibitem[Frail1997]{}
Frail, D. et al. 1997, Nature 389, 261

\bibitem[Frail2002]{}
Frail, D. A. \& Berger, e. 2002, GCN Circular 1418

\bibitem[Fryer, Woosley, \& Hartmann (1999)]{Fry99b}
Fryer, C. L., Woosley, S. E., Hartmann, D. H. 1999, ApJ, 526, 152 

\bibitem[Gandolf2002]{}
Gandolfi, G. et al. 2002, {\it BeppoSAX Results on Short Gamma-Ray
Bursts}, in WH2001{$^{**}$}.

\bibitem[Graziani94]{}
Graziani, C., Lamb, D. Q. 1994, in AIP Conf. Proc. 307, Gamma-Ray Bursts,
ed. G. J. Fishman, J. J. Brainerd, and K. Hurley (New York: AIP), 227

\bibitem[Hurley(1992)]{}
Hurley, K. 1992, in AIP Conf. Proc. 265, Gamma-Ray Bursts,
ed. W. Paciesas \& G. Fishman (New York: AIP), 3

\bibitem[Hurley(2002a)]{}
Hurley, K. 2002a, ApJ, 567, 447

\bibitem[Hurley(2002b)]{gcn1402}
Hurley, K., et al. 2002b, GCN Circ. 1402

\bibitem[Hurley(2002c)]{gcn1402}
Hurley, K., et al. 2002c, GCN Circ. 1407

\bibitem[Kawai(2002)]{WH2001_WXM}
Kawai, N., et al. 2002, {\it In-Orbit Performance of the WXM 
Instrument on HETE}, in WH2001{$^{**}$}.

\bibitem[Kehoe2001]{}
Kehoe, R. et al.  2001, Ap.J. 554, L159

\bibitem[Kouveliotou1993]{}
Kouveliotou, C. et al. 1993, ApJ, 413, L101

\bibitem[Kulkarni20020]{}
Lamb, D. Q. 2002, {\it Gamma-Ray Bursts as a Probe of Cosmology}, in
WH2001{$^{**}$}.

\bibitem[Lamb1993]{}
Lamb, D. Q., Graziani, C. \& Smith, I. A. 1993, ApJ, 413, L11

\bibitem[Lamb2002]{}
Lamb, D. Q. et al. 2002, GCN Circular 1403

\bibitem[MacFadyen1999]{}
MacFadyen, A. \& Woosley, S. 1999, ApJ, 524, 262

\bibitem[Miceli2002]{}
Miceli, A., Lamb, D. Q., Zucker, D., Covey, K., Dembicky, J. \&
Hastings, N. C. 2002, GCN Circular 1416

\bibitem[Monnelly2002]{}
Monnelly, G. et al. 2002, {\it HETE Soft X-Ray Camera Imaging: 
Calibration, Performance, and Sensitivity}, in WH2001{$^{**}$}.

\bibitem[Paciesas2001]{}
Paciesas, W. S., Preece, R. D., Briggs, M S., and Malozzi, R. S. 2001, 
in Gamma-Ray Bursts in the Afterglow Era, (Rome, Italy, 17-20 October
2000), ESO Astrophysics Symposia, Springer (Berlin ), p. 13

\bibitem[Paradijs1997]{}
van Paradijs, J. et al. 1997, Nature 386, 686

\bibitem[Park2002]{}
Park, H. S., Williams, G. G., Lindsay, K. 2002, GCN Circular 1404

\bibitem[Ricker(2002a)]{gcn1399}
Ricker, G. R., et al. 2002, GCN Circ. 1399

\bibitem[Ricker(2002`b)]{WH2001_HETE} Ricker, G.R., et al. 2002, {\it
High Energy Transient Explorer (HETE): Mission and Science Overview},
in WH2001{$^{**}$}.

\bibitem[Ruffert1999]{}
Ruffert, M. \& Janka, H. 1999, A\&A, 344, 573

\bibitem[Schmidt2001]{}
Schmidt, M. 2001, ApJ, 559, L79

\bibitem[Tavani1998]{}
Tavani, M. 1998, ApJ, 497, L21

\bibitem[Vietri \& Stella (1998)]{Vie98}
Vietri, M., \& Stella, L. 1998, ApJL, 507, L45 

\bibitem[Vietri \& Stella (1999)]{Vie99}
Vietri, M., \& Stella, L. 1999, ApJL, 527, L43

\bibitem[West2002]{}
West, D. 2002, GCn Circular 1406


\end{thebibliography}
\end{document}